\documentclass[11pt,twoside]{article}
\usepackage{asp2014}
\pdfoutput=1
\aspSuppressVolSlug
\resetcounters
\begin{document}

\title{Automated Spectral Classification of Galaxies using Machine Learning Approach on Alibaba Cloud AI platform (PAI)}
% Note the position of the comma between the author name and the 
% affiliation number.
% Author names should be separated by commas.
% The final author should be preceded by "and".
% Affiliations should not be repeated across multiple \affil commands. If several
% authors share an affiliation this should be in a single \affil which can then
% be referenced for several author names.
% See ManuscriptInstructions.pdf and ASPmanual2010.pdf 3.1.4 for more details
\author{Yihan Tao,$^1$ Yanxia Zhang,$^1$Chenzhou Cui$^1$, Ge Zhang$^2$ 
\affil{$^1$National Astronomical Observatories, Chinese Academy of Sciences, Beijing, China; \email{y.tao@nao.cas.cn}}
\affil{$^2$Alibaba Cloud Computing, Beijing, China}}

% This section is for ADS Processing.There must be one line per author.
\paperauthor{Yihan Tao}{y.tao@nao.cas.cn}{}{National Astronomical Observatories, Chinese Academy of Sciences}{}{Beijing}{Beijing}{100012}{China}
\paperauthor{Yanxia Zhang}{zyx@bao.ac.cn}{}{National Astronomical Observatories, Chinese Academy of Sciences}{}{Beijing}{Beijing}{100012}{China}
\paperauthor{Chenzhou Cui}{ccz@bao.ac.cn}{}{National Astronomical Observatories, Chinese Academy of Sciences}{}{Beijing}{Beijing}{100012}{China}
\paperauthor{Ge Zhang}{13911031680@139.com}{}{Alibaba Cloud Computing}{}{Beijing}{Beijing}{}{China}

\begin{abstract}
Automated spectral classification is an active research area in astronomy at the age of data explosion. While new generation of sky survey telescopes (e.g. LAMOST and SDSS) produce huge amount of spectra, automated spectral classification is highly required to replace the current model fitting approach with human intervention. Galaxies, and especially active galactic nucleus (AGNs), are important targets of sky survey programs. Efficient and automated methods for galaxy spectra classification is the basis of systematic study on physical properties and evolution of galaxies. To address the problem, in this paper we carry out an experiment on Alibaba Cloud AI plaform (PAI)\footnote{https://data.aliyun.com/product/learn?spm=5176.8142029.388261.302.9bXCPD} to explore automated galaxy spectral classification using machine learning approach. Supervised machine learning algorithms (Logistic Regression, Random Forest and Linear SVM) were performed on a dataset consist of ~ 10000 galaxy spectra of SDSS DR14, and the classification results of which are compared and discussed. These galaxy spectra each has a subclass tag (i.e. AGNs, Starburst, Starforming, and etc.) that we use as training labels.
\end{abstract}

\section{Introduction}
 With large scale sky survey telescopes, e.g. SDSS, LAMOST etc., spectra are acquired at PB or even TB level per minutes. Confronting high volume of spectra, classification is the first step for astronomers to carry out research on various types of objects on the sky. Currently, both SDSS and LAMOST data release use model matching method to divide the spectra into star, galaxy, quasar and unknown categories. More specifically, in SDSS, stars and galaxy are further classified into subclass. As such survey goes along, astronomers will get much more spectra data. Meanwhile, automated classification of spectra into subclass with less human intervention are highly in need. 

\section{Classification of Galaxy Spectra}
In this paper, we focus on the classification of galaxy spectra as automatically classifying emission line galaxies is complex and it believed to be meaningful for understanding galaxy formation and evolution. Machine learning approach has been applied to enable automated classification of galaxy spectra. \citet{2002SPIE.4847..371Z} \citet{2015MNRAS.451..629S,2015MNRAS.453..122S} applied Support Vector Machine (SVM) and Artificial Neural Networks (ANN) to a series of line ratio features of galaxy spectra provided by MPA/JHU, and classify galaxies as AGN, composite or SFG. Emission line galaxies are usually using Baldwin-Phillips-Terlevich (BPT) diagram, the features of which are line ratios. These works acquire highly acceptable classification accuracy but the problem is measure the emission-line ratios from original spectra needs to be done accurately by astronomers in advance. Therefore, we are interested in automated extraction of features from original spectra. Specifically, in this poster paper, we address the following research questions:How can classification of AGNs be done through automated machine learning algorithms using cloud computing platform and services? How does each algorithms perform in galaxy spectra classification?
\section{Data}
\label{sec:data}
The dataset for this experiment consists of over 10000 galaxy spectra from SDSS DR14. These spectra were labeled according to their subclass field. AGN were coded as 1, and the rest types including STARFORMING, STARBURST and No Tags were coded as 0. Eighty percent of the dataset were split as training set and the rest were used as test set.

To get the common window of rest-frame spectra and retain the majority of emission lines, the spectra wavelength were truncated to 2642 pixels from 3700-6800 $\overset{\circ}{A}$. 

Feature selection is an important aspect of machine learning. As the spectra usually contains thousands of pixels, dimension reduction based on appropriate feature representation is a valuable research question. Principle Components Analysis (PCA) is a technique to extract feature and compress the spectra, which is commonly used in spectra classification
\citep{1998MNRAS.298..361B,1998MNRAS.295..312S,2010AJ....139.1261M,2013hell.confQ..24K,2014IAUS..306..301B}. PCA with 20 components were applied to the original flux and used as features in classification.

\section{Experiments on PAI}
Alibaba PAI is a platform designed for artificial intelligence which incorporate a series of data preprocessing, feature engineering and commonly used machine learning algorithms. In this experiment, we employed 3 commonly used algorithms, namely logistic regression, random forest and linear support vector machine (SVM), to explore their effectiveness in automated classification of galaxy spectra on a selected SDSS dataset as described in section  \ref{sec:data}.

\articlefigure{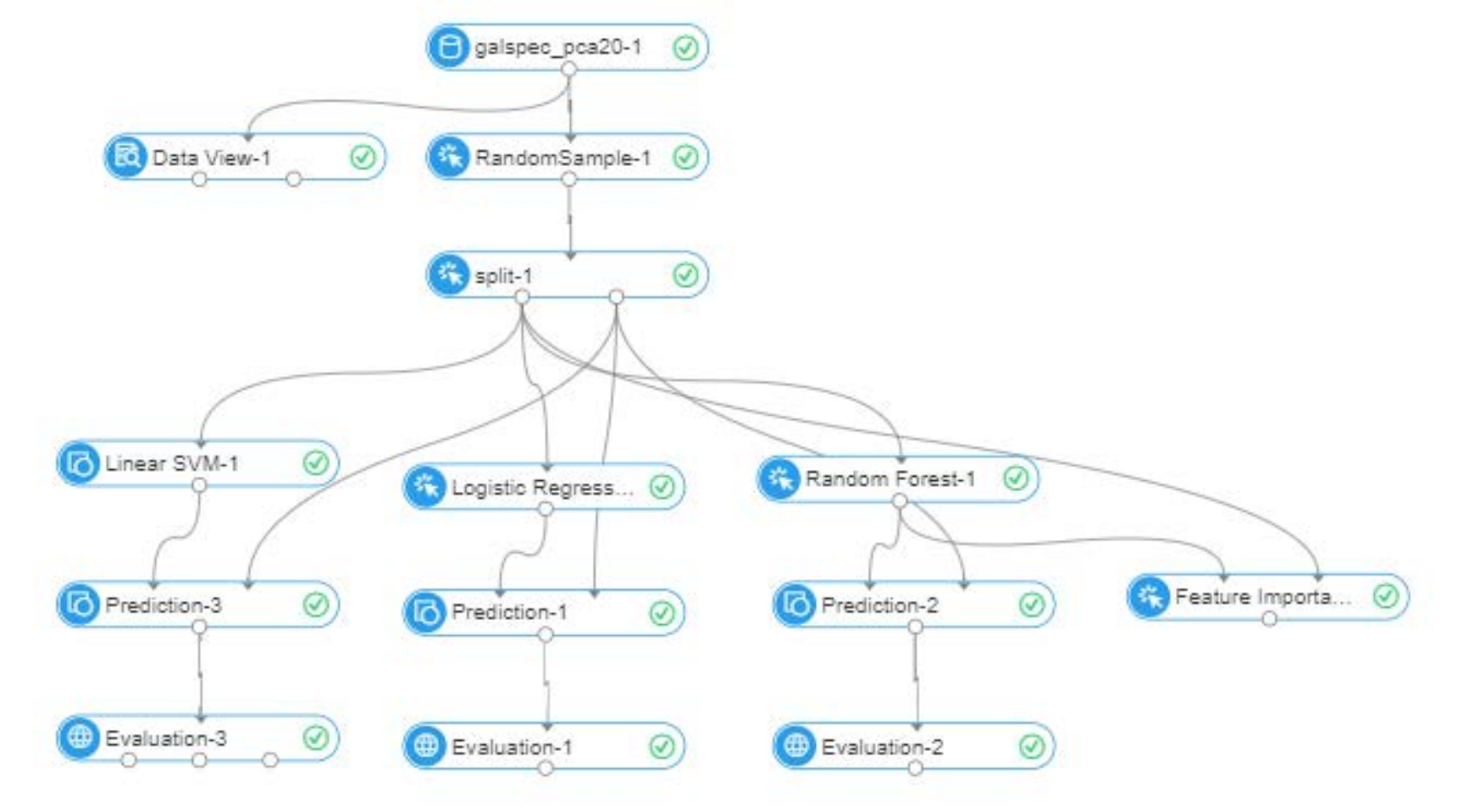}{fig:Experiment_Flow}{The overall process of the experiment on PAI}

The overall flow of the experiment on PAI is shown in Figure \ref{fig:Experiment_Flow}. The experiment start from generating 10000 random samples from around 10600 pre-processed spectra data.

\section{Results and Discussion}
In this section, we present the results of the experiment and compare the performance of the three different algorithms in terms of precision, recall and F1-score. The efficiency of algorithms is not presented as the execution time for each algorithms are almost the same. This might because that our dataset is relatively small and PAI is powerful to produce results in a few minutes. In the following, we discuss the classification accuracy and feature importance in more detail.

\subsection{Classification Results}
 As Table \ref{tab:results} shows, all three classification algorithms (i.e linear regression, linear SVM and random forest) with 20 components PCA reached higher than 92\% on precision and recall. More specifically, the three methods have nearly the same F1-score. This might because that the limitation of classification algorithms lies in the feature representation, i.e. linear PCA.

\begin{table}[!h]
\caption{Comparison of classification results}
\label{tab:results}
\smallskip
\begin{center}
{\small
\scalebox{0.7}{ 
\begin{tabular}{lccc}  % l = left, c = centered
\tableline
\noalign{\smallskip}
 & Linear Regression/ with feature elimination & Linear SVM/ with feature elimination & Random Forest/ with feature elimination \\
\noalign{\smallskip}
\tableline
\noalign{\smallskip}
Precision & 0.9293/ 0.9293
 & 0.9227/0.9289
 & 0.9215/0.9255 \\
Recall & 0.9283/ 0.9283 & 0.9212/0.9272
 &0.9215/0.9255\\
F1-score & 0.9284/ 0.9284
 & 0.9214/0.9274 &0.9215/0.9255 \\
\noalign{\smallskip}
\tableline
\end{tabular}
}}
\end{center}
\end{table}

\subsection{Feature Importance}

\articlefiguretwo{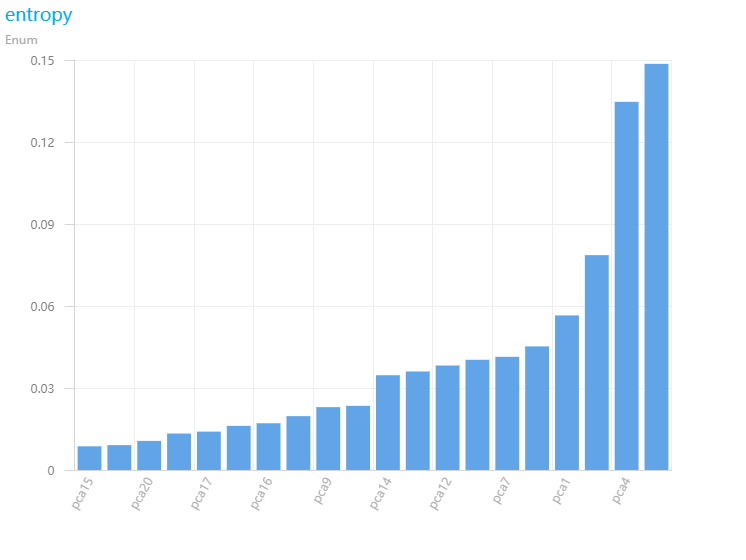}{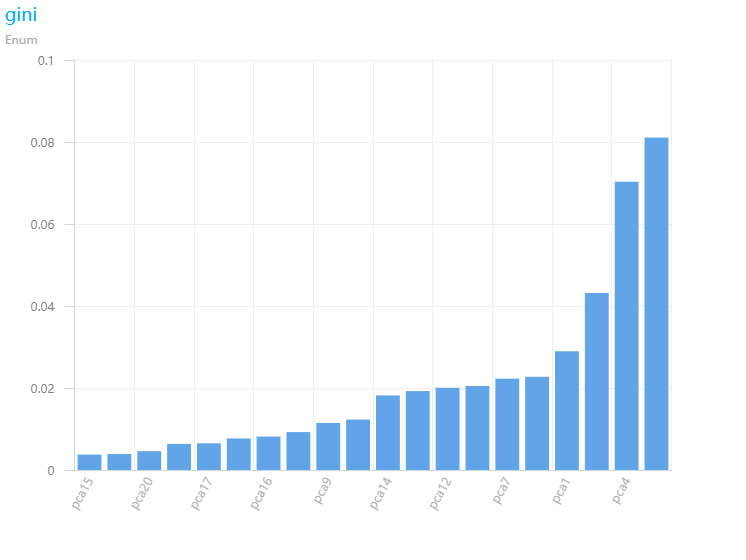}{Feature_Importance}{Feature importance based on entropy}

In this experiment, one of the algorithm we adopted is random forest, an ensembled decision tree algorithm to classify the galaxy spectra. During the training, the random forest algorithm also calculated feature importance. As shown in Figure \ref{Feature_Importance}

By selecting the top 10 important features (feature elimination), we acquire even better precision and recall score. This indicate that we could achieve better results with most important features extracted.

\section{Conclusions and Future Work}
In this experiment, we simply apply PCA with 20 components and it acquires over 92\% accuracy in classification of AGNS.
However, the spectra can not be present well simply in linear combination of spectral features. A better extraction and representation of features might lead to better classification results. The future plan is to investigate other ways, such as  deep neural networks (auto encoder, RBM etc.), to extract features and comparing with the results of using PCA.

Also, an limitation of PAI is that it does not support random search or grid search functions, we adjust the hyper parameters manually and by experience. In the future, we should perform a complete search in the parameter space.
\acknowledgements
This work is supported by the Young Researcher Grant of National Astronomical Observatories, Chinese Academy of science, National Natural Science Foundation of China (NSFC)(11503051, 61402325) and the Joint Research Fund in Astronomy (U1531111, U1531115, U1531246, U1731125, U1731243) under cooperative agreement between the NSFC and Chinese Academy of Sciences (CAS). Data resources are supported by Chinese Astronomical Data Center (CADC) and Chinese Virtual Observatory (China-VO).

\bibliography{adass17}

\end{document}